# Switching magnetic spin-states using small magnetic fields in compositionally complex Sm(M7)O$_3$

R. K. Dokala[1*], M. Geers[2], P. Nordblad[1], R. Clulow[2*], and R. Mathieu[1*]

[1]*Department of Materials Science and Engineering, Uppsala University, Box 35, 751 03, Uppsala, Sweden*
[2] *Department of Chemistry - Ångström laboratory, Uppsala University, Box 538, 751 21, Uppsala, Sweden*

**Abstract**

High-entropy perovskites (HEPs) offer a unique platform for exploring magnetic phenomena arising from extreme *B*-site chemical disorder. In Sm(M7)O$_3$, where there are 7 cations in equal amounts at the *B*-site; M = Ti, Cr, Mn, Fe, Co, Ni, Cu), we observe long-range antiferromagnetic ordering near 105 K accompanied by a small but robust excess magnetic moment intrinsic to the chemically disordered lattice. This uncompensated moment is evident from ZFC–FC irreversibility, shifts in the isothermal M(H) loops, and discrete remanent states identified through direct-current–demagnetization measurements. Remarkably, cooling fields as small as ±20 Oe are sufficient to select the direction of the excess moment, and the chosen magnetic state remains stable against applied fields up to 50 kOe. A low-temperature anomaly in the remanent magnetization further reveals a secondary contribution from the Sm$^{3+}$ sublattice, although the primary origin of the excess moment resides in the *B*-site AFM sublattice.

[*]Corresponding authors:

ravikirandokalaiitg@gmail.com
rebecca.clulow@kemi.uu.se
roland.mathieu@angstrom.uu.se

## 1. Introduction

High-entropy oxides (HEOs) have recently attracted considerable interest as platforms where extreme chemical disorder produces emergent magnetic states that do not occur in conventional ordered lattices.[1-3] In several perovskite HEOs, long-range antiferromagnetic or ferrimagnet-like order survives despite the random arrangement of multiple transition-metal cations, while the underlying disorder frequently generates ill-compensated AFM with small residual moments embedded in an AFM lattice.[4-6] This behavior mirrors phenomena long known in classical dilute AFMs such as $FeF_2$ and $Fe_xZn_{1-x}F_2$, where random-field and random-exchange effects lead to uncompensated local moments, excess remanence, and apparent exchange-bias–like shifts even in the absence of ferromagnetic interfaces.[7-9] The $FeF_2$ literature demonstrates that such uncompensated moments can be stabilized by local anisotropy and disorder, giving rise to magnetic remanent states and field-history-dependent magnetization.[10] Similar signatures have recently been reported in compositionally complex perovskites—including the high-entropy perovskite family—where AFM order coexists with small, frozen-in moments arising from nanoscale exchange imbalance.[2] Recent advances further show that these disorder-borne moments can be manipulated with surprisingly small cooling fields. In chemically complex fluorides and oxides, cooling fields of only a few tens of Oe can reorient pinned AFM spinstates, induce exchange-bias-like offsets, and generate large low-field magnetic switching effects.[11] In high-entropy perovskites (HEPs), such sensitivity is particularly notable because the disordered $B$-site network creates a rich landscape of magnetic spin configurations, as highlighted in the recent literature on magnetically ordered HEPs.[2,4] In this work, we investigate the high-entropy perovskite $Sm(M_7)O_3$, where equal fractions of seven transition-metal ions populate the $B$-site. We show that the system develops an antiferromagnetic order with a small but robust excess magnetic moment intrinsic to the high-entropy lattice. Most notably, this excess moment can be reversibly switched using cooling fields as small as ±20 Oe, and at low temperatures, the selected magnetic state remains stable up to high applied fields. These results demonstrate that high-entropy perovskites offer a powerful route for low-field tuning of uncompensated antiferromagnetic states and suggest that such behavior may be a general feature of magnetically ordered HEOs independent of the rare-earth ion occupying the $A$-site.

## 2. Experimental methods

$SmTi_xCr_xMn_xFe_xCo_xNi_xCu_xO_3$, denoted $Sm(M7)O_3$ in the following where $x = 1/7$, was synthesised using conventional solid-state synthesis with a final sintering temperature of 1150 °C for 48 h in air. $Sm_2O_3$ (Alfa Aesar, >99.9 %) was pre-dried at 975 °C overnight prior to weighing and combined in stoichiometric proportions with the metal oxides, $TiO_2$ (Merck), $Cr_2O_3$ (Alfa Aesar), $MnO_2$ (Cerac), $Fe_2O_3$ (Alfa Aesar), $Co_3O_4$, NiO (Sigma Aldrich) and CuO (Cerac), all with purity >99.5 %. Co (Höganäs, >99%) was oxidised to $Co_3O_4$ at 600 °C for 6 h, and its purity confirmed by powder X-ray diffraction prior to synthesis. The reagents were ground with minimum (approx. 2 mL) ethanol for 30 min. and pressed into a pellet. The pellet was heated in air using an alumina crucible at 900 °C for 12 h, 1000 °C for 24 h, 1100 °C for 48 h and 1150 °C for 48 h with intermediate grindings. A heating rate of 5 °C/min and cooling rate of 10 °C/min were used during sintering. Powder X-ray diffraction (PXRD) patterns were collected at room temperature using a Bruker D8 ADVANCE diffractometer equipped with TWIN/TWIN setup (40 kV, 40 mA) using Cu K$\alpha$ radiation in Bragg-Brentano geometry. Data were collected using a Lynx-eye XE-T position sensitive detector, operated over an angular $2\theta$ range of 10-80° with a step size of 0.02 corresponding to a Q range of 0.75 – 4.95 Å$^{-1}$. Rietveld refinements of the structural model were completed using the FullProf program.[12] The sample composition was obtained using a ZEISS Leo 1550 field emission scanning electron microscope (SEM) equipped with an AZtec energy dispersive X-ray detector for spectroscopy analysis (EDS). Pellets of the sintered powder were attached to conducting carbon tape and point EDS measurements were performed using an accelerating voltage of 20 kV on 9 spots (see Fig. SM1)[13]. The magnetic properties, including the temperature-dependent zero field-cooled (ZFC), field-cooled (FC) magnetization, and ac susceptibility were acquired using a superconducting quantum interference design (SQUID) measurement system from Quantum Design Inc. Magnetic field ($H$) dependent magnetization ($M$) measurements were recorded on the same equipment at a constant temperature. M(H) hysteresis curves were recorded at different temperatures. The direct current demagnetization (DCD) was recorded by cooling the sample in a large magnetic field (-50 kOe) and applying and removing a reverse (positive) magnetic field of increasing amplitude.[14] The remanence value after each reverse field pulse is plotted as a function of the reverse field.

## 3. Results and Discussions

Figure 1 presents the powder x-ray diffraction pattern of the high-entropy perovskite Sm(M7)O$_3$, together with the corresponding Rietveld refinement. The results show that the compound crystallizes with a distorted perovskite structure, adopting orthorhombic *Pnma* symmetry ($a$ = 5.5797(7) Å, $b$ = 7.6515(7) Å, $c$ = 5.3778(5) Å). The calculated profile provides a good fit to the observed data aside from one very weak reflection present at 2.49 Å$^{-1}$ which arises from trace amounts of an unknown impurity. The high quality of the fit and observed phase purity suggest that the Sm has been fully incorporated into the A site of the perovskite. The transition metals are randomly distributed over the octahedral *B*-sites and the Sm cations are located on the 12-fold oxygen coordination *A*-sites. The result of EDS point analysis of 9 spots indicates an average cation composition of Sm$_{51(1)}$Ti$_{9.7(9)}$Cr$_{7.4(5)}$Mn$_{7(1)}$Fe$_{7.5(5)}$Co$_{7.2(5)}$Ni$_{5(1)}$Cu$_{3.0(5)}$. This is close to the expected ratio for the *B*-site metals of 7.14. On average, the sample is deficient in Cu, most likely as a result of the low melting point of CuO. The structural model highlights the extensive chemical disorder on the *B*-site, where seven distinct transition-metal cations occupy the octahedral framework in a completely random manner. This high configurational complexity—characteristic of high-entropy oxides—is expected to produce significant local distortions and competing exchange pathways, which play a central role in the magnetic behavior discussed in later sections. Figure 2a shows the temperature dependence of the magnetization measured under an applied field of $H_{DC}$ = 50 Oe following zero-field-cooled (ZFC) and field-cooled (FC) protocols, after resetting the magnet to ensure near zero cooling field. Both ZFC and FC curves exhibit a clear onset of magnetic ordering near $T \approx 105$ K. Below this temperature, a pronounced bifurcation between the ZFC and FC branches is observed, with the FC magnetization remaining significantly larger than the ZFC magnetization over the entire low-temperature range. The magnetization decreases below ~ 10 K. The in-phase component of the ac susceptibility, $\chi_{ac}$, measured at frequencies of 1.7, 17, and 170 Hz, is shown in Fig. 2b. A prominent feature is observed near $T \approx 105$ K, a sharp maximum of the magnetic transition coinciding with the magnetic transition seen in the dc magnetization data. Within the measured frequency range, the position of this feature shows negligible frequency dependence. Figure 3a shows the temperature dependence of the magnetization measured under an applied field of $H_{meas}$ = 1 kOe after cooling the sample in small fields of +20 Oe and −20 Oe. For $H_{CF}$ = +20 Oe, the magnetization curve remains positive over the entire temperature range. For $H_{CF}$ = −20 Oe, the magnetization is negative at low temperatures and the two branches converge near the magnetic ordering temperature (≈ 105 K) for both cooling conditions. Figure 3b displays isothermal field-dependent, M(H) magnetization curves measured under ZFC conditions at 2, 25, and 75 K. At 2 K, the magnetization varies approximately linearly with the applied magnetic field over the entire measured range up to ± 50 kOe. As the temperature increases coercivity and hysteresis is appearing. The slope of the M(H) curves decreases with increasing temperature as well. In the inset, a zoomed view of the M(H) curve at 2K is shown, which in fact is not symmetrical across the origin and rather seems shifted up on the magnetization scale or to the negative side on the field axis. The observation together with the positive ZFC magnetization in Fig. 2a suggests a small positive background field in the measurement system which plays the same role as the +20 Oe cooling field in Fig. 3a. Figure 3c presents the temperature-dependent magnetization measured at a higher applied field of $H_{meas}$ = 50 kOe (instead of 1 kOe) after cooling the sample in small opposite fields of -20 Oe and + 20 Oe. The magnetization curves remain similar at high temperatures but the one cooled in -20 Oe deviates from the one cooled in +20 Oe below 25 K. The magnetization values at low temperatures, thus depend on the sign of the cooling field. The results shown in Fig. 3c suggests that the magnetic susceptibility of the material include three contributions: $\chi_{Sm}$ the paramagnetic susceptibility of the $Sm^{3+}$ sublattice, $\chi_{3d}$ which represents the field- and temperature-dependent susceptibility of the $B$-site cations responsible for the AFM order near ~105 K, and $M_0$, an excess moment arising from the ill-compensated sublattices of the AFM structure as:

$$M(H,T)/H = \chi_{Sm}(H,T) + \chi_{3d}(H,T) + M_0(H,T)/H.$$

The direction of the excess moments being determined by the sign of the cooling field. Figure 3d presents field-dependent magnetization curves measured at 2 K under FC conditions for several cooling fields ranging from −10 kOe to +50 kOe. All M(H) curves exhibit approximately linear field dependence up to ±50 kOe. The slopes of the curves are similar for different cooling fields, but the curves are vertically or horizontally shifted relative to each other depending on the magnitude and sign of the cooling field. Interestingly, the shift is relatively little dependent on the strength of the field but highly sensitive to the sign of the cooling field. Figure 3e shows a magnified view of the low-field region of the M(H) curves at 2 K. A small hysteresis is visible, and the measurements taken under ZFC* conditions (which is an FC at −0.3 Oe to compensate the background field and achieve near-zero effective cooling field as confirmed by series of FC measurements, see Fig. SM3) closely follow the central branch, indicating minimal offset compared to curves obtained after cooling in larger positive or negative fields. M(H) at higher temperatures (as shown in Fig. 3b and Fig. SM6) shows a significant coercivity, which is not in reach at low temperature. It is noteworthy that the ΔM (= M(*after cooling in +20 Oe*)-M(*after cooling in -20 Oe*)) at 50 kOe measured from M(H) curves, as well as the ΔM from the MT curves after cooling in +20 Oe and -20 Oe and measured under 50 kOe are same.

To probe the switching of the magnetization after field cooling, we have performed a series of direct-current–demagnetization (DCD) measurements. DCD has extensively been used to probe the magnetization switching of magnetic materials and is used here to investigate the stability of the magnetic configurations under increasing magnetic fields. In this protocol, the sample is cooled from the paramagnetic state to the target temperature in a strong negative field (−50 kOe), the field is removed at 2 K, and the magnetization is subsequently recorded upon applying and removing reverse fields until +50 kOe for each temperature. The remnant magnetization, $M_{REM}$ after each reverse field pulse is recorded and plotted as a function of the reverse field. The resulting $M_{REM}(H_{rev})$ curves, shown in Fig. 4, display a characteristic two-step evolution at all temperatures between 2 and 40 K. The magnetization initially remains on a low-moment plateau and then undergoes an abrupt increase at a well-defined switching field. Notably, the magnitude of this switching field coincides with the critical fields $H_{Critical}(T)$ extracted from the maxima in dM/dH obtained under ZFC conditions as shown in Fig. 4d. The temperature dependence of the remnant magnetization $M_{REM}(T)$, plotted in Fig. 4c, provides additional insight as the zero-field values of $M_{REM}$ exhibit a pronounced anomaly below ~10 K which suggests the Sm contribution to the magnetic signal (cf. the M(T) curves at low T in Fig. 2).

The present results demonstrate that the antiferromagnetic order established in Sm(M7)O$_3$ near 105 K is accompanied by a small but robust excess magnetic moment. This uncompensated moment manifests through ZFC–FC irreversibility, vertical shifts of the M(H) curves after field cooling, and discrete remanent states observed in DCD remanence measurements. Such behavior is inconsistent with a fully compensated collinear antiferromagnet [15] and instead indicates an intrinsic ill compensation within the ordered AFM structure. Given the statistically distributed multicomponent *B*-site, this excess moment is most naturally attributed to local imbalance of AFM sublattices in each grain of the polycrystalline sample arising from chemical disorder and competing exchange interactions inherent to high-entropy perovskites. A salient feature of this uncompensated moment is its strong sensitivity to the cooling field. Remarkably, cooling fields as small as ±20 Oe are sufficient to select the sign of the magnetization, and the selected state remains stable even under applied fields as large as 50 kOe. This behavior indicates that the excess moment is associated with magnetic configurations separated by sizable anisotropy barriers, such that a weak symmetry-breaking field can bias the system into one of two energetically equivalent but oppositely oriented states.[16] Once selected, these states exhibit high stability, pointing to an effective anisotropy that pins the uncompensated component of the antiferromagnetic structure. At low temperatures, an additional contribution from the Sm$^{3+}$ sublattice becomes evident. The anomaly observed below ~10 K in the remanent magnetization, in this temperature range, suggests that the Sm moments couples antiferromagnetically to the excess moments. However, the primary origin of the excess moment and its field tunability resides in the *B*-site–driven AFM framework rather than in the RE contribution. Importantly, these observations are not expected to be unique to Sm-based compounds. The emergence of a controllable uncompensated moment appears to be a generic consequence of an AFM order in chemically disordered HEP, independent of the specific RE ion occupying the *A*-site.

## 4. Conclusions

In this work, we report the ability to reversibly tune and stabilize distinct magnetic states using fields as small as ±20 Oe in a compositionally complex perovskite with seven *B*-site cations Sm(M7)O$_3$. We speculate that this is an inherent feature of HEPs, owing to the intrinsically ill compensated AFM structure in those systems.

These results thus indicate that this low field tuning may be a generic and technologically relevant feature of magnetically ordered HEPs.

**Acknowledgements**

We thank the Olle Engkvist Stiftelse (Grant No. 224-0046) for financial support. R.C and M.G acknowledge financial support from the Olle Engkvist Stiftelse (Grant No. 229-0390) and the Åforsk foundation (Grant No. 24-349).

[1] S. Schweidler, M. Botros, F. Strauss, Q. Wang, Y. Ma, L. Velasco, G. C. Marques, A. Sarkar, C. Kübel, H. Hahn, J. A. Hagmann, T. Brezesinski, B. Breitung, "High-entropy materials for energy and electronic applications," Nat Rev Mater **9**, 266 (2024).
[2] R. Witte, A. Sarkar, R. Kruk, B. Eggert, R. A. Brand, H. Wende, H. Hahn, "High-entropy oxides: An emerging prospect for magnetic rare-earth transition metal perovskites," Phys. Rev. Materials **3**, 034406 (2019).
[3] C. M. Rost, E. Sachet, T. Borman, A. Moballegh, E. C. Dickey, D. Hou, J. L. Jones, S. Curtarolo, J. P.Maria, "Entropy-stabilized oxides," Nat. Commun. **6**, 8485 (2015).
[4] J. Cedervall, R. Clulow, H. L. B. Boström, D. C. Joshi, M. S. Andersson, R. Mathieu, P. Beran, R. I. Smith, J. C. Tseng, M. Sahlberg, P. Berastegui, S. Shafeie, "Phase stability and structural transitions in compositionally complex $LnMO_3$ perovskites," J. Solid State Chem. **300**, 122212 (2021).
[5] A. Sarkar, Q. Wang, A. Schiele, M. R. Chellali, S. S. Bhattacharya, D. Wang, T. Brezesinski, H. Hahn, L. Velasco, B. Breitung, "High-Entropy Oxides: Fundamental Aspects and Electrochemical Properties," Adv. Mater. **31**, 1806236 (2019).
[6] P. Pramanik, R. Clulow, D. C. Joshi, A. Stolpe, P. Berastegui, M. Sahlberg, R. Mathieu, "Spin glass states in multicomponent layered perovskites," Sci. Reps. **14**, 3382 (2024).
[7] D. C. Joshi, P. Nordblad, R. Mathieu, "Ferromagnetic excess moments and apparent exchange bias in $FeF_2$ single crystals," Sci. Rep. **9**, 18884 (2019).
[8] D. C. Joshi, P. Nordblad, R. Mathieu, "Random fields and apparent exchange bias in the dilute Ising antiferromagnet $Fe_{0.6}Zn_{0.4}F_2$," Sci. Rep. **10**, 14588 (2020).
[9] E. Maniv, R. A. Murphy, S. C. Haley, S. Doyle, C. John, A. Maniv, S. K. Ramakrishna, Y. L. Tang, P. Ercius, R. Ramesh, A. P. Reyes, J. R. Long, J. G. Analytis, "Exchange bias due to coupling between coexisting antiferromagnetic and spin-glass orders," Nat. Phys. **17**, 525, (2021).
[10] M. J. Benitez, O. Petracic, H. Tuysuz, F. Schuth, H. Zabel, "Fingerprinting the magnetic behavior of antiferromagnetic nanostructures using remanent magnetization curves," Phys. Rev. B **83**, 134424 (2011).
[11] Z. Deng, X. Wang, M. Wang, F. Shen, J. Zhang, Y. Chen, H. L. Feng, J. Xu, Y. Peng, W. Li, J. Zhao, X. Wang, M. Valvidares, S. Francoual, O. Leupold, Z. Hu, L. H. Tjeng, M. R. Li, M. Croft, Y. Zhang, E. Liu, L. He, F. Hu, J. Sun, M. Greenblatt, C. Jin, "Giant Exchange-Bias-Like Effect at Low Cooling Fields Induced by Pinned Magnetic Domains in $Y_2NiIrO_6$ Double Perovskite," Adv. Mater. **35**, 2209759 (2023).
[12] J. Rodríguez-Carvajal, "Recent advances in magnetic structure determination by neutron powder diffraction," Physica B: Condens. Matter **192**, 55 (1993).
[13] See Supplementary Materials for EDS data and more magnetometry results.
[14] P. I. Mayo, K. O'Grady, R. W. Chantrell, J. A. Cambridge, I. L. Sanders, T. Yogi, and J. K. Howard, "Magnetic measurement of interaction effects in CoNiCr and CoPtCr thin film media," J. Mag. Mag. Mat. **95**, 109-117 (1991).
[15] J. M. D. Coey, "Non-collinear spin arrangement in ultra-fine ferrimagnetic crystallites," Phys. Rev. Lett. **27**, 1140 (1971).
[16] J. Nogués and I. K. Schuller, "Exchange bias," J. Magn. Magn. Mater. **192**, 203 (1999).

**Figures**

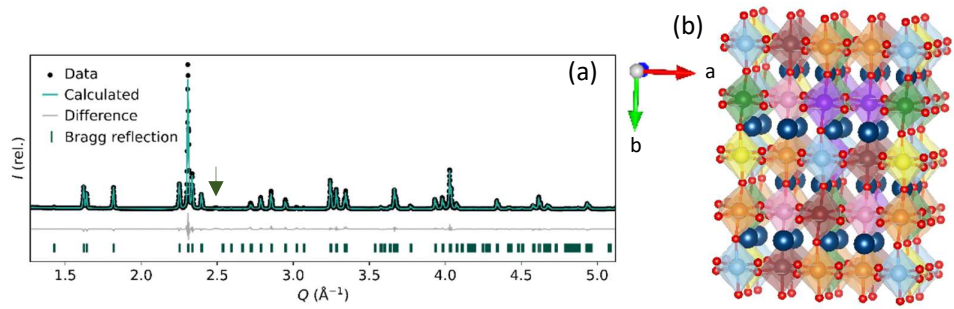

**Fig 1**. (a) Rietveld refinements (green line) of powder X-ray diffraction data (*black circles*), with Bragg reflections (*green tick marks*). $R_{wp}$ = 7.97 %, $\chi^2$ = 60.0. One weak reflection arising from an unidentified impurity is identified with an arrow. (b) the crystal structure of Sm(M7)O$_3$ derived from Rietveld analysis, with randomly distributed *B*-site (*multi-coloured octahedra*), Sm = *dark blue*, O = *red*.

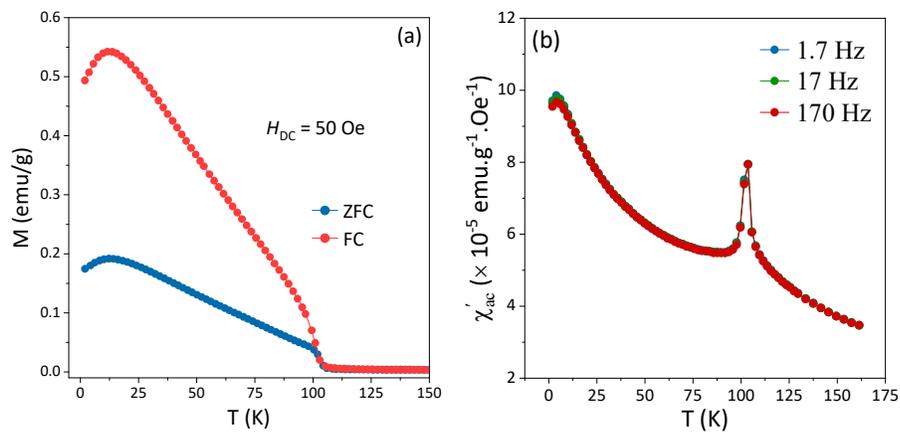

**Fig 2**. (a) Temperature dependent magnetization in an applied field of $H_{DC}$ = 50 Oe under ZFC and FC protocols after resetting the magnet to ensure the cooling field to be near zero ZFC and FC. (b) In-phase component of *ac*-susceptibility (*ac* amplitude $h_{ac}$ = 4 Oe).

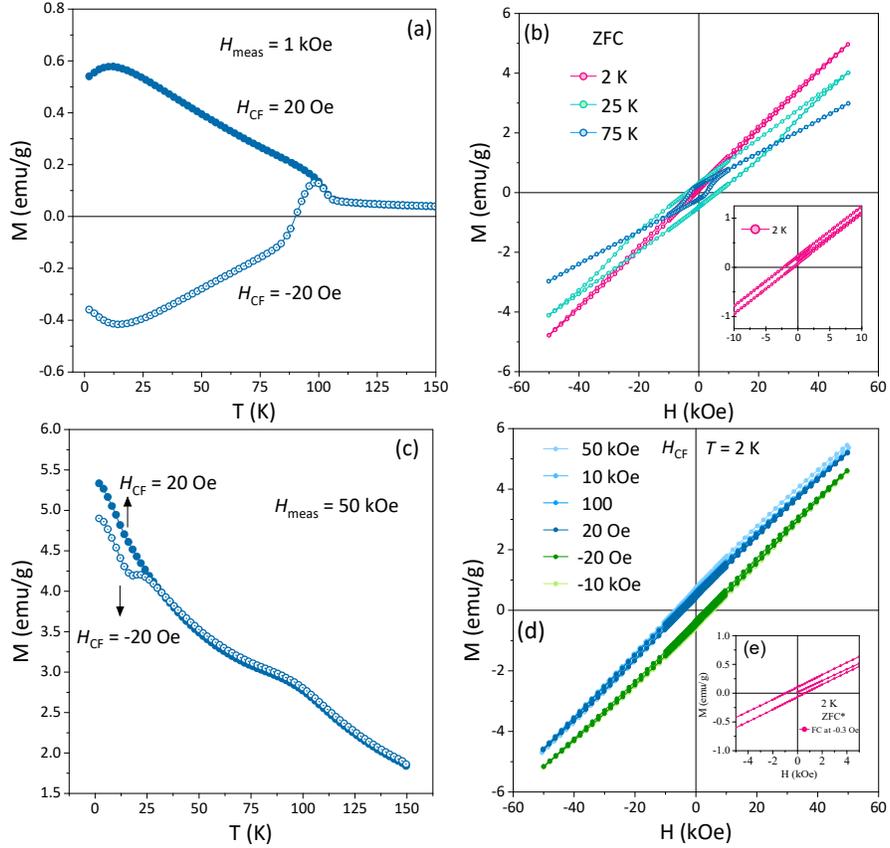

**Fig 3**. Magnetization curve recorded versus temperature after cooling in small field of -20 Oe or 20 Oe (a) $H_{meas}$ = 1 kOe (*See Fig. SM4 for complete "ZFC"/FC curves*), and (c) $H_{meas}$ = 50 kOe (*See Fig. SM5a for complete "ZFC"/FC curves*) (b) Field dependent magnetization curves at 2 K, 25 K, and 75 K. *Inset* shows the zoomed version of the ZFC M(H) loop at 2K. (d) Magnetization curves recorded versus field after cooling in the same fields as well as fields of larger magnitude (see legends: positive and negative cooling fields are plotted using shades of blue and green, respectively), (e) M(H) recorded in ZFC* condition where, ZFC* (FC at -0.3 Oe) indicates that the small background correction was made to obtain field as close as possible to zero field.

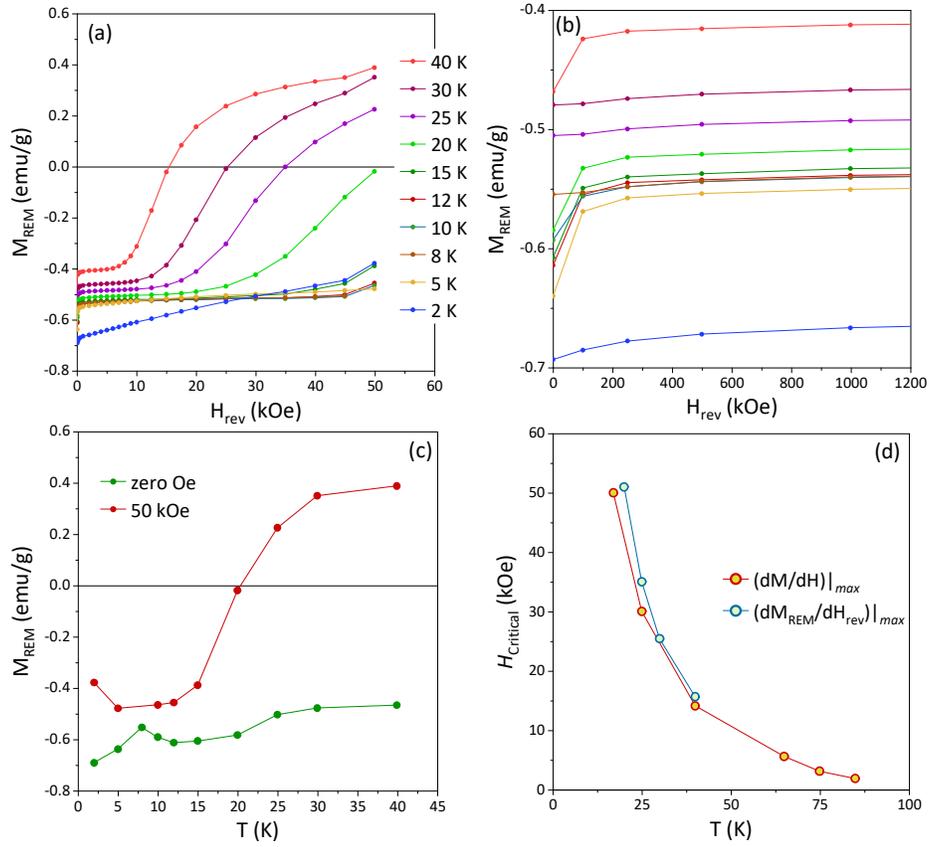

**Fig 4**. (a) DCD remanence curves obtained after cooling in -50 kOe and applying and removing reverse magnetic field of increasing magnitude from zero to +50 kOe. (b) Zoomed view of (a). (c) The value of the remanence in zero and after a 50 kOe field pulse is plotted with respect to temperature. (d)Temperature dependence of $H_{Critical}$ extracted from $(dM/dH)_{max}$: the maximum slope in M(H) curves shown in Fig. SM6, and $(dM_{REM}/dH_{rev})_{max}$ identified as the switching field for which $M_{REM}$ is zero (*i.e.* switches from negative to positive) in (a).